\documentclass[aps,prl,floatfix,twocolumn,showpacs,superscriptaddress,longbibliography]{revtex4-1}
\usepackage{graphics}
\usepackage{epsfig}
\usepackage{amsmath}
\usepackage{amssymb}
\usepackage[usenames]{color}
\usepackage[caption=false]{subfig}
\usepackage{float}
\usepackage{hyperref}

\begin{document}
\newcommand{\beq}{\begin{equation}}
\newcommand{\eeq}{\end{equation}}
\newcommand{\mean}[1]{\ensuremath{\left\langle #1 \right\rangle}} 

\title{Disordered auxetic networks with no re-entrant polygons}

\author{Varda F. Hagh}
\affiliation{Department of Physics, Arizona State University, Tempe, AZ 85287-1504, USA}

\author{M. F. Thorpe}
\affiliation{Department of Physics, Arizona State University, Tempe, AZ 85287-1504, USA}
\affiliation{Rudolf Peierls Centre for Theoretical Physics, University of Oxford, 1 Keble Rd, Oxford OX1 3NP, England}

\date{\today}

\begin{abstract}
It is widely assumed that disordered auxetic structures (i.e. structures with a negative Poisson's ratio) must contain re-entrant polygons in $2$D and re-entrant polyhedra in $3$D. Here we show how to design disordered networks in $2$D with {\it{only}} convex polygons. The design principles used allow for any Poisson ratio $-1 < \nu < 1/3$ to be obtained with a prescriptive algorithm. By starting from a Delaunay triangulation with a mean coordination $\mean{z} \simeq 6$ and $\nu \simeq 0.33$ and removing those edges that decrease the shear modulus the least, without creating any re-entrant polygons, the system evolves monotonically towards the isostatic point with $\mean{z} \simeq 4$ and $\nu \simeq -1$.
\end{abstract}


\maketitle

Consider a homogeneous extension of a rod whose sides are free. If we apply a uniform force at the two ends of the rod in opposite directions, it will undergo a transverse expansion when compressed and a transverse compression when stretched along the applied forces. This is the familiar behavior of most materials. This deformation can be quantified by Poisson's ratio, which is defined as the negative ratio of transverse contraction strain to longitudinal expansion strain. 
In $d$ dimensions, the Poisson's ratio of any bulk material is related to its bulk ($K$) and shear ($G$) elastic moduli by~\cite{milton1992composite}:

\begin{equation}
\label{eqn:poisson}
\nu = \frac {dK-2G}{d(d-1)K+2G}
\end{equation}
which reduces to $\nu = (K - G)/(K + G)$ in $2$D. Since for any material $K, G \geq 0$ for stability, we must have:

\begin{equation}
\label{eqn:range}
(K = 0) \quad  -1 \leq \nu \leq \frac{1}{d-1}   \quad(G = 0)
\end{equation}
where $\nu = (d-1)^{-1}$ corresponds to an incompressible fluid or rubber with a vanishingly small shear modulus compared to its bulk modulus. Note that $\nu = 0$ corresponds to $K - 2G/d =\lambda = 0$ where $\lambda$ is the Lame' constant~\cite{salenccon2012handbook}, as occurs in cork for example~\cite{greaves2013poisson}. Thus a negative $\nu$ corresponds to negative Lame' constant which is not forbidden by thermodynamics but is unusual. 
{\it{Normal}} materials have a positive Poisson's ratio. From a continuum elasticity point of view, this is because most materials have a larger resistance to changes in their volume (described by the bulk modulus $K$) compared to resistance to changes in their shape (defined by their shear modulus $G$)~\cite{prawoto2012seeing}. 

Eq.~(\ref{eqn:poisson}) suggests that by designing a structure where $K < 2G/d$ or simply $K < G$ in $2$D, one can fabricate materials with a negative Poisson's ratio. These types of materials and structures are called auxetic. 
The concept of a negative Poisson's ratio goes back to Saint-Venant in 1848~\cite{Venant} for anisotropic materials. In the modern era, this concept was extensively described by Love in 1944~\cite{love2013treatise}, and later was investigated by Gibson in 1982~\cite{gibson1982mechanics}. In 1987, Lakes re-fabricated conventional polymer foams with a positive Poisson's ratio by heating under pressure to create re-entrant structures on the sub-millimeter scale which then led to foams with a negative Poisson's ratio that were isotropic~\cite{lakes1987foam}. These investigations suggested that auxetic behavior is the result of a mechanism that involves the geometrical structure of the material and its deformation under compressive load. A variety of of these materials were designed and fabricated at the end of 80's and the beginning of 90's~\cite{,burns1987negative, caddock1989microporous, evans1991design, alderson1992fabrication, alderson1994auxetic}. Since then, many similar efforts (theoretically, computationally and experimentally) have led to auxetic materials~\cite{yang2004review}. These include auxetic cellular foams~\cite{warren1990negative, choi1995nonlinear, evans1994auxetic, chan1997fabrication,gibson1999cellular, smith2000novel, evans2000auxetic, grima2006negative}, auxetic regular and disordered networks~\cite{ prall1997properties, evans1991molecular, keskar1992negative,baughman1993crystalline, grima2006auxetic, blumenfeld2012theory, cabras2014auxetic, hanifpour2017mechanics, reid2018auxetic}, microporous polymers~\cite{caddock1989microporous, pickles1995effect, pickles1996effects, alderson1997interpretation}, and laminated fiber composites~\cite{hine1993modelling,ravirala2006negative}. In this paper, we will focus on disordered auxetic networks~\cite{reid2018auxetic} with only convex and no re-entrant polygons. 

\begin{figure}[htp]
\centering
\includegraphics[width=6cm,height=2.8cm]{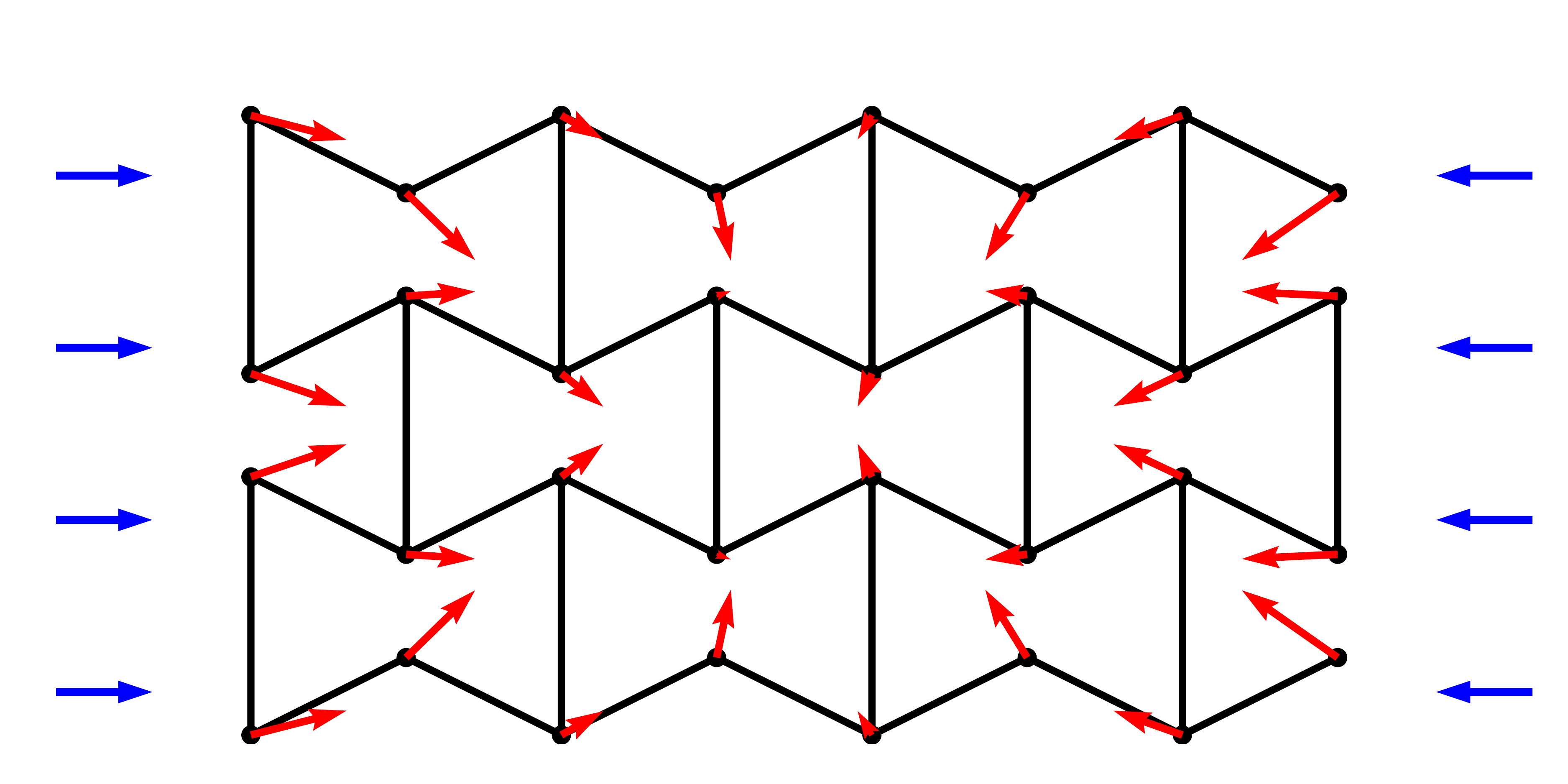}
\caption{ (color online) A hexagonal re-entrant honeycomb with bow tie shaped polygons. This type of re-entrance is common in engineered materials with negative Poisson's ratio. The horizontal blue arrows on the sides represent the external load that is applied to the system. The red arrows, attached to the nodes, show the movements of all the nodes in response to the external load. The magnitudes of arrows have been multiplied by $10^3$ to make them visible to the eye.}
\label{fig:reentrant}
\end{figure}

It is important to note that theoretical studies of auxetic materials fall into {\it{two}} distinct classes. In the {\it{first}} category, of interest here, the material is over-constrained with all the elastic moduli being non-zero and proportional to the spring constant(s) in the system where for simplicity, we assume the same spring constant for all edges present. In the {\it{second}} category of auxetic structures, the structure is under-constrained and a single internal {\it{mechanism}} or {\it{floppy mode}} is involved in which the associated eigenvector shows auxetic behavior but there is no restoring force and all the elastic moduli are zero (for a recent treatment with references see~\cite{borcea2015geometric, borcea2017periodic, sun2012surface}). All the edges retain their original lengths when the system undergoes a deformation.  In this case, Eq.~(\ref{eqn:poisson}) cannot be used for the Poisson ratio as $K=G=0$, and instead the ratio of traverse to longitudinal strain is used. Note that for almost any material with a few floppy modes (few meaning between say 2 and 5), a negative Poisson's ratio can usually be achieved by using a well-chosen linear combination of floppy mode eigenstates. Thus we regard the {\it{first}} category as being more challenging and focus on that here as it is of the most interest for experimental fabrication. 

Most presently known auxetics with non-zero elastic constants and nearest neighbor central forces are networks with a re-entrant node structure~\cite{lakes1993advances}. A re-entrant or pointed node in a network is a node where two adjacent edges make an angle greater than $180^{\circ}$. A classic example of this can be seen in Figure~\ref{fig:reentrant}. The mechanism of deformation for these types of networks is very well understood and involves the collapse of all the bow tie units as they are pushed from any direction. 

In this paper, we demonstrate a computational method to build two dimensional disordered networks with Poisson's ratios in the range $-1<\nu<1/3$ and {\it{convex polygons only}}. We have been unable to find any examples of \emph{disordered} networks in the literature with controllable Poisson's ratios and nearest neighbor central forces that did not contain re-entrant polygons. The known auxetic structures such as chiral honeycombs~\cite{prall1997properties} that do not possess any re-entrance, have unit cells with a specific type of symmetry (e.g. rotational, chiral, mirror, etc.). When there is such a symmetry in the system, a single mechanism like unrolling can drive the system auxetic. Our interest on the other hand, is in linear elasticity where the edges of a disordered network are springs and the network is over-constrained. Such a network has no symmetry (other than the repetitive structure associated with the supercell) and when all the polygons are convex, its structure resembles that of glassy and jammed networks that are widely studied in rigidity theory~\cite{thorpe1983continuous, ellenbroek2015rigidity}.


\begin{figure}[htp]
\centering
\subfloat[]{\includegraphics[width=7cm]{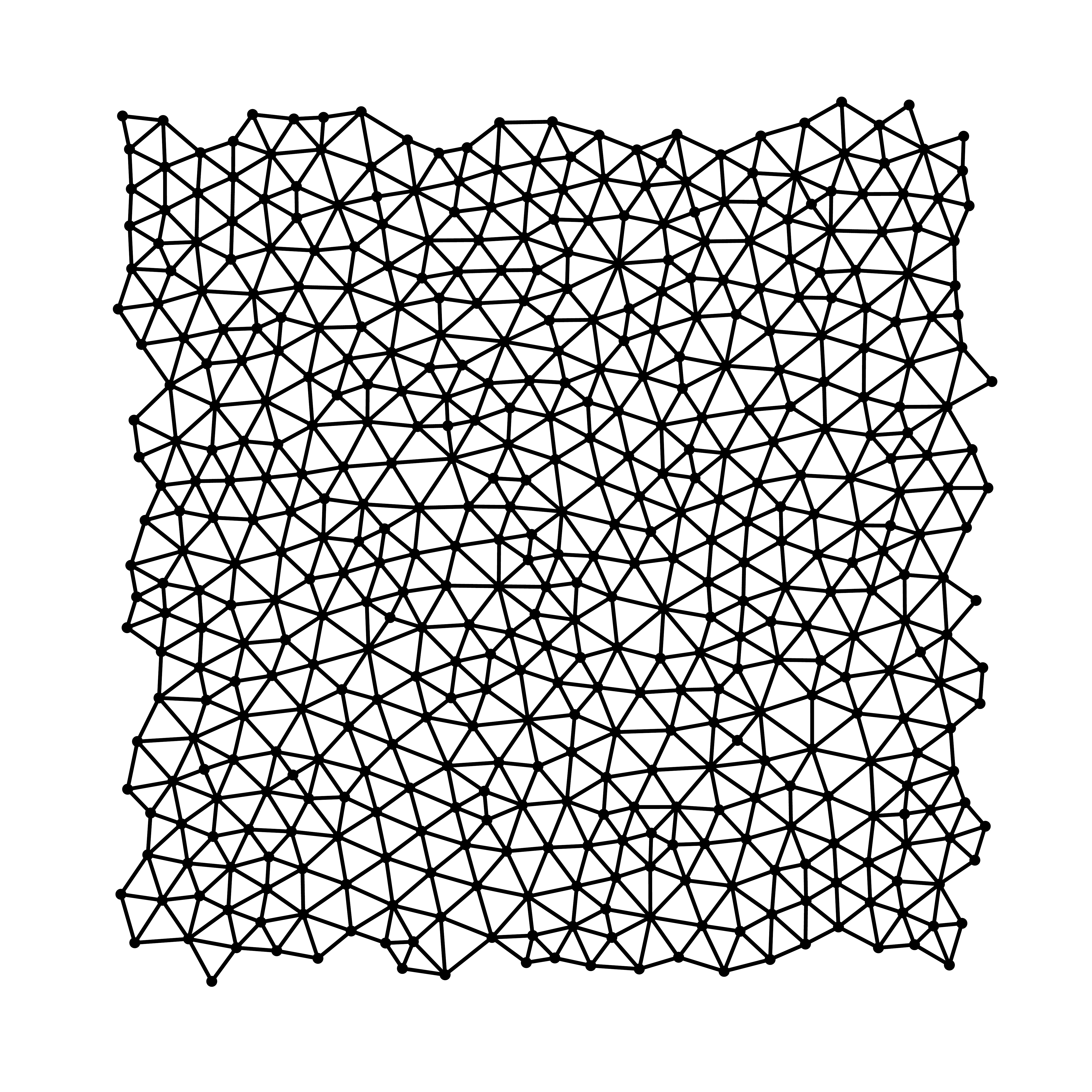}}\\[-2ex] 

\subfloat[]{\includegraphics[width=7cm]{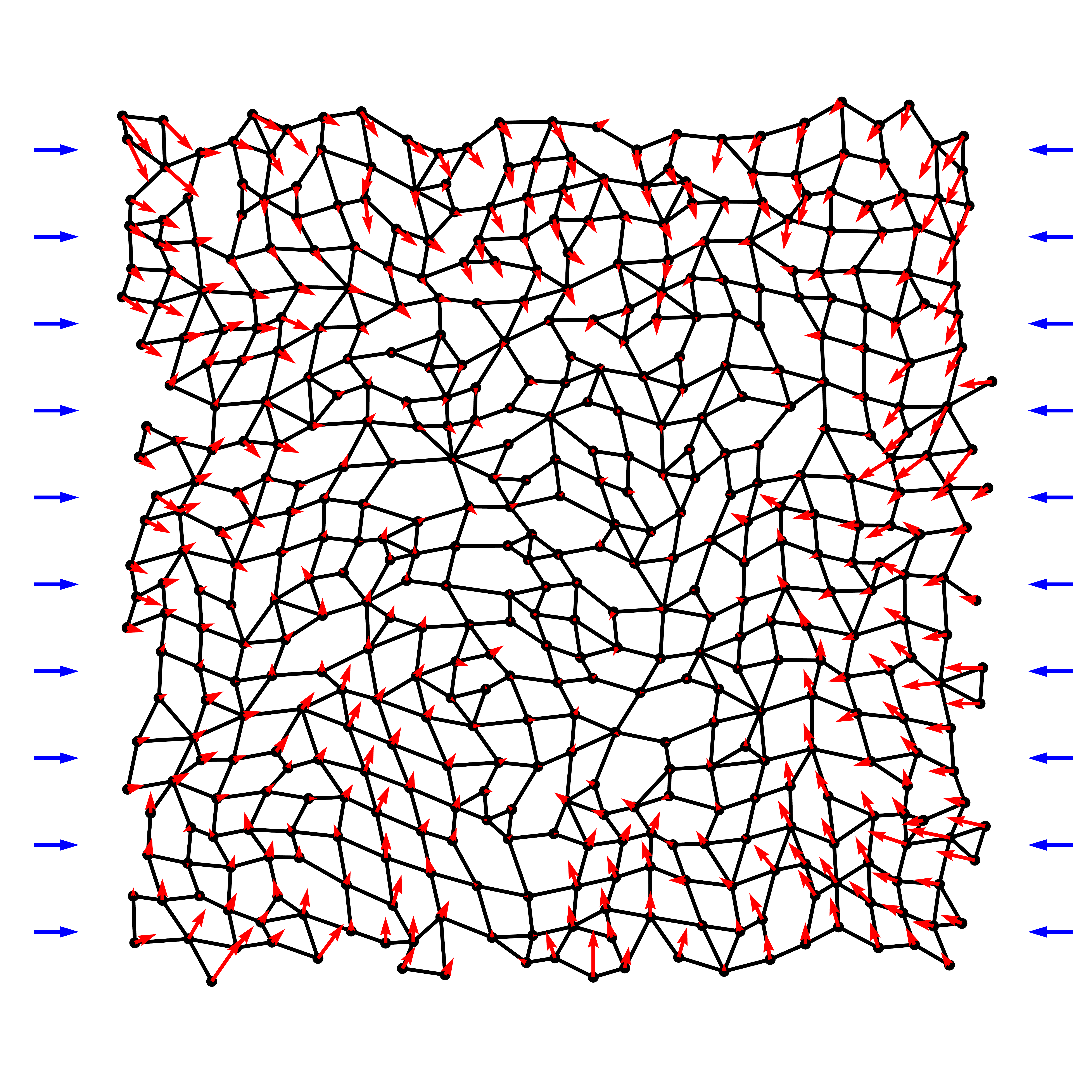}}
\caption{(color online) a) A disordered triangular network with mean coordination  $\mean{z} = 6$ before removing any edges. b) The same network after removing one third of the edges while the convexity of all the polygons is conserved. The mean coordination number is $\mean{z} \simeq 4$ and the network has a negative Poisson's ratio of $\nu = -0.9998$. The horizontal blue arrows on the sides represent the external load that is applied to the system. The red arrows, attached to the nodes, show the movements of all the nodes $u_i$ in response to the external load. The magnitudes of arrows have been multiplied by $10^3$ to make them visible to the eye.}
\label{fig:network}
\end{figure}

In the recent years, topological optimization methods have been widely used to design networks with specific elastic and mechanical properties~\cite{goodrich2015principle, reid2018auxetic, hagh2018jamming}. In this paper, we use tuning by pruning method to generate networks that have a finite shear modulus of order $1$ and an infinitesimal bulk modulus of order $O(1/N)$, so that $K \ll G$. Here, $N$ denotes the number of nodes in the network. In the limit $N \rightarrow \infty$, the bulk modulus of these systems becomes zero and therefore the Poisson's ratio, as defined by Eq.~(\ref{eqn:poisson}), becomes exactly $\nu = -1$. The networks are generated by starting from a fully triangulated spring network with mean coordination $\mean{z} = 6$ and periodic boundary conditions. The starting network is a Delaunay triangulation~\cite{lee1980two} of a set of points generated by Poisson disk sampling in $2$D~\cite{bridson2007fast,dunbar2006spatial}. An example can be seen in Figure \ref{fig:network}-a.

The contribution of different edges to the elastic moduli of a harmonic spring network can span over several orders of magnitude~\cite{hexner2017linking}; affecting the bulk and shear moduli in very different ways in some cases~\cite{goodrich2015principle, hexner2018role}. This means removing some of the edges can cause a significant drop in either the value of bulk or shear modulus (or both), while the removal of some other edges does not change the moduli by a significant amount. The wide distribution of edge response in these networks allows us to identify and remove those edges that have the minimum contribution to the changes in bulk or shear modulus. For example, removing edges that have the minimum contribution to the bulk modulus can be used to build networks with a finite bulk modulus that resemble a jammed system~\cite{hagh2018jamming}.

Here, we remove those edges that have a smaller contribution to the shear modulus of the system. The shear modulus is measured by compressing the network in the horizontal direction, while stretching it in the vertical direction. This deformation causes a change in the lengths of the springs which all are assumed to have a unit spring constant, $k=1$ (N/m). These springs have no physical width and there is no energy associated with bending them. Therefore the effective spring constant is only based on stretching or compressing the edges. The energy stored in the system ($E$) is then measured and the shear modulus is calculated using the following equation~\cite{ellenbroek2009non}:

\begin{equation}
\label{eqn:shear_mod}
G = \frac{1}{2} \frac{E}{A \delta^2}
\end{equation}
where $A$ denotes the total area of the network and $\delta$ is the strain applied to the system. Note that the shear modulus $G$ is independent of $\delta$ in the linear regime ($\delta \ll 1$). If we iteratively remove the edges with the smallest contribution to $G$ from a mean coordination $\mean{z} = 6$ down to $\mean{z} \simeq 4$, the Poisson's ratio will monotonically go from $\nu \simeq 1/3$ to $\nu \simeq -1$ and the resulting network will have a larger resistance to shearing than to hydrostatic compression. This method of pruning naturally introduces re-entrance into the system. To avoid the emergence of re-entrant nodes and hence maintain the convexity of all the polygons in the network, one more crucial condition is added to the pruning protocol. This extra condition is Hilbert's mechanical stability~\cite{lopez2013jamming}, which is imposed on all nodes at each step of the pruning process. The Hilbert's condition guarantees that each node must have at least $d+1$ incident edges and the geometrical arrangement of edges is such that applied forces can cancel each other out. This geometrical condition is useful for the global mechanical rigidity of the network. In $2$D this means no angles between adjacent edges can be greater than $180^{\circ}$, and hence re-entrance is prevented. 

As an aside, all the removed edges could be replaced with very weak springs (for example a thousand times weaker) to get back to the original Delaunay triangulation which would still be auxetic. The only polygons then are triangles which of course are convex. However, this illustrates that to be meaningful, the notion of convexity has to be tied in with springs of comparable magnitude. Any auxetic network with non-convex polygons, can also be modified by adding a single auxiliary node inside each re-entrant polygon and connecting that new node to the nodes of the polygon with very weak springs. This will form a local triangulation and will make the network entirely convex. But, again, it is not a meaningful way of circumventing the meaning of {\it{convex}}.

To be precise, we first generate disordered triangular networks with mean coordination $\mean{z} = 6$ that are Delaunay triangulation of a Poisson disk sampling with $N=500$ points on a $2$D plane. We then loop over all the edges and collect those that will not violate the Hilbert's stability condition if removed. This guarantees that the removal of an edge will not create any non-convex polygons in the network. The contribution of each removable edge to the shear modulus of the system is then measured and the edge list is sorted in an ascending order based on the value of their contribution. Finally, we select the first 10\% of the edges with smallest contributions to $G$ and remove one of these randomly. More than one third of the edges need to be removed to drive the original triangulated network to an auxetic network with $\nu \simeq -1$ and the mentioned process is repeated at each step. We could have selected the edges with smallest contribution, but chose one out of the smallest 10\% to demonstrate that the result is robust, and the results are virtually identical.

Figure~\ref{fig:network}-b shows an auxetic network generated by this method. As can be seen from the figure, there are no re-entrant nodes introduced to the system, and yet the network has a negative Poisson's ratio $\nu = -0.9998$. The small deviation of the Poisson's ratio from $-1$ is a finite size effect and would vanish in the $N \rightarrow \infty$ limit. The red arrows show the displacements $\{u_i\}$ of all the nodes when we apply a small strain of order $\delta = 10^{-4}$ in the horizontal direction (shown by the blue arrows) and let the system relax. The center of mass has been fixed here which leads to $\sum_{i=1}^{N} u_i = 0$ and therefore there is not much motion happening at the central parts of the network. The scale of these motions are magnified $10^3$ times to make them visible to the eye, since we are in the linear regime and the motions are infinitesimal. However this magnification is for visualization only as anharmonic effects are present at such large displacements for non-collinear networks of harmonic springs.

The mechanism behind the auxetic behavior of the networks built here is not trivial, and a simple explanation has eluded us, but lies within the method used to build them. The generating process is very cooperative, as in each step an edge is removed based on how its contribution to the shear modulus is compared to all the other edges in the network. This cooperative process adds to the complexity of the mechanism that involves the deformation of such systems, and is a particular example of a larger phenomena that involves pruning spring networks in special ways to obtain desired properties. Another example is to produce jammed networks~\cite{hagh2018jamming}, and yet another to produce allosteric effects of a similar kind to those seen in proteins~\cite{rocks2017designing}. To date these are all empirical algorithms and the underlying mathematics remains to be understood. 

\begin{figure}[H]
\centering
\includegraphics[width=7cm]{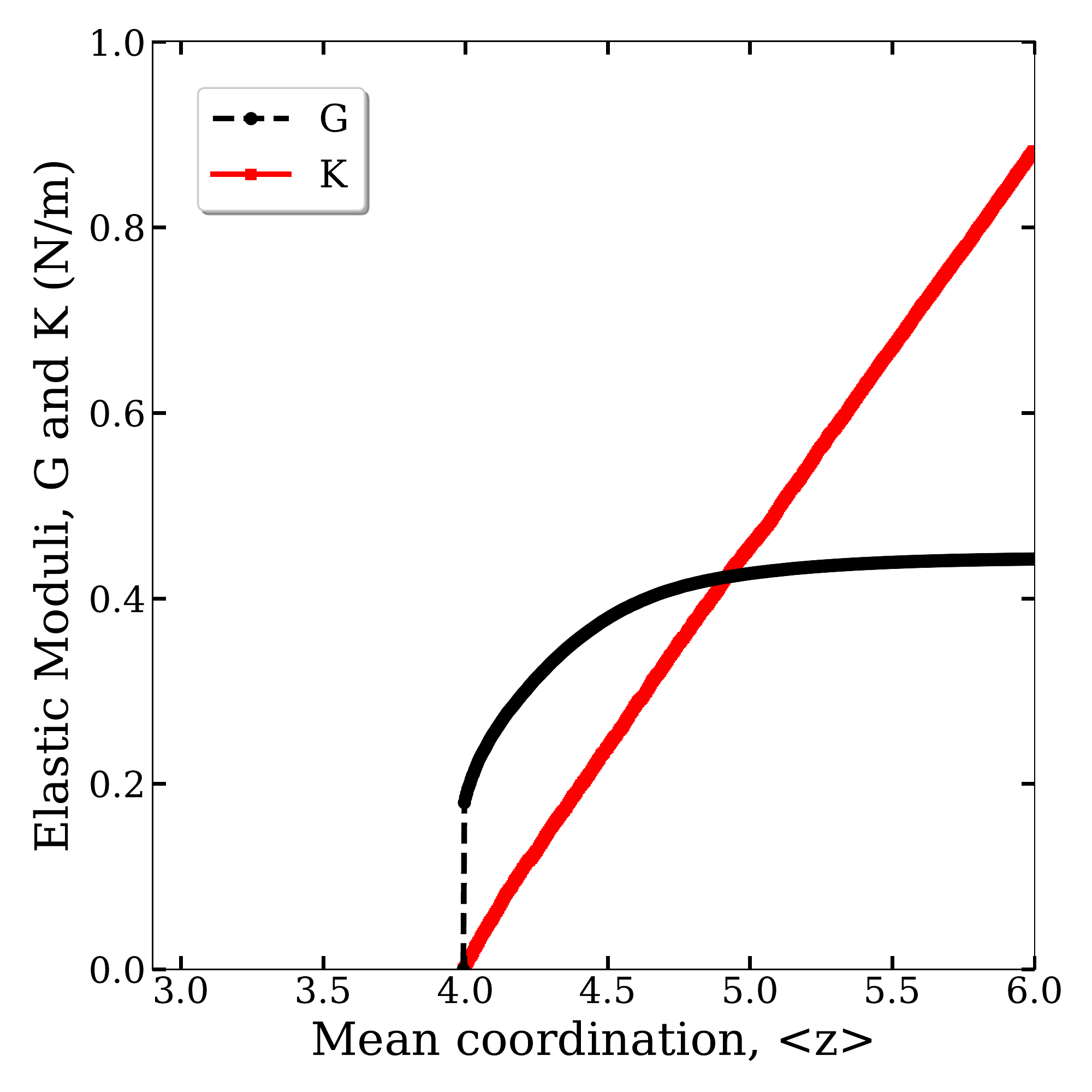}
\caption{(color online) The plots of the shear ($G$) and bulk ($K$) moduli as the edges that minimally affect $G$ are pruned from mean coordination $\mean{z} = 6$ down to $\mean{z} \simeq 4$. The results are ensemble averaged over 50 samples, each with $N=500$ nodes.}
\label{fig:plots}
\end{figure}

We monitor both the shear and bulk moduli of these networks as they are pruned. The bulk modulus is measured in a similar way to the shear modulus by using Eq.~(\ref{eqn:shear_mod}). Figure~\ref{fig:plots} shows the behavior of both bulk and shear moduli against the mean coordination of the system. The mean coordination is defined as the average number of edges at each node. All data points are ensemble averaged over 50 samples with $N=500$ nodes. Both these elastic moduli decrease monotonically as the edges are removed~\cite{lord1945theory}, as required by general principles.

At the starting point, the bulk modulus of a triangular network is greater than the value of its shear modulus; therefore the Poisson's ratio is a positive number, as can be seen from Eq.~(\ref{eqn:poisson}). It should be noted that if the nodes in a network are connected by central forces and if {\it{every node}} is a center of symmetry, then because of the Cauchy condition between elastic constants, $c_{12} = c_{44}$, the Poisson's ratio would be $\nu =(d+1)^{-1}$ which in $2$D gives $\nu=1/3$. This is the case for a $2$D regular triangular network~\cite{feng1985effective} and is also closely true for a Delaunay triangulation of the kind shown in Figure~\ref{fig:network}-a. The algorithm used here to select the removed edges aims to keep the shear modulus of the system above zero. Since the changes in bulk modulus are only loosely correlated with changes in the shear modulus, driving the network to maximize the shear modulus does not force it to also maximize the bulk modulus. The bulk modulus decreases linearly as it would do with random dilution~\cite{ellenbroek2015rigidity}. This makes the difference between bulk and shear become smaller and smaller until at about $\mean{z} \simeq 4.92$ they become equal. For any edges removed after this, the shear modulus is larger than the bulk modulus and therefore the Poisson's ratio becomes negative. As $\mean{z} = 4$ is approached, the bulk modulus goes to zero while the shear modulus remains non-zero; therefore the Poisson's ratio approaches $-1$. Note that the last edge that is removed, takes the system to the isostatic point plus one edge~\cite{hagh2018jamming} where there is one state of self-stress in the system~\cite{sun2012surface} and the shear modulus is of order $1$, while the bulk modulus is $O(1/N)$. The removal of an additional edge is meaningless as this would take the system to the isostatic point where the total number of degrees of freedom and constraints are balanced such that the only remaining floppy modes are the macroscopic rigid motions. At the isostatic point, the network is still mechanically stable but both the bulk and shear moduli are exactly zero and therefore Poisson's ratio becomes undefined.

\begin{figure}[htp]
\centering
\includegraphics[width=7cm]{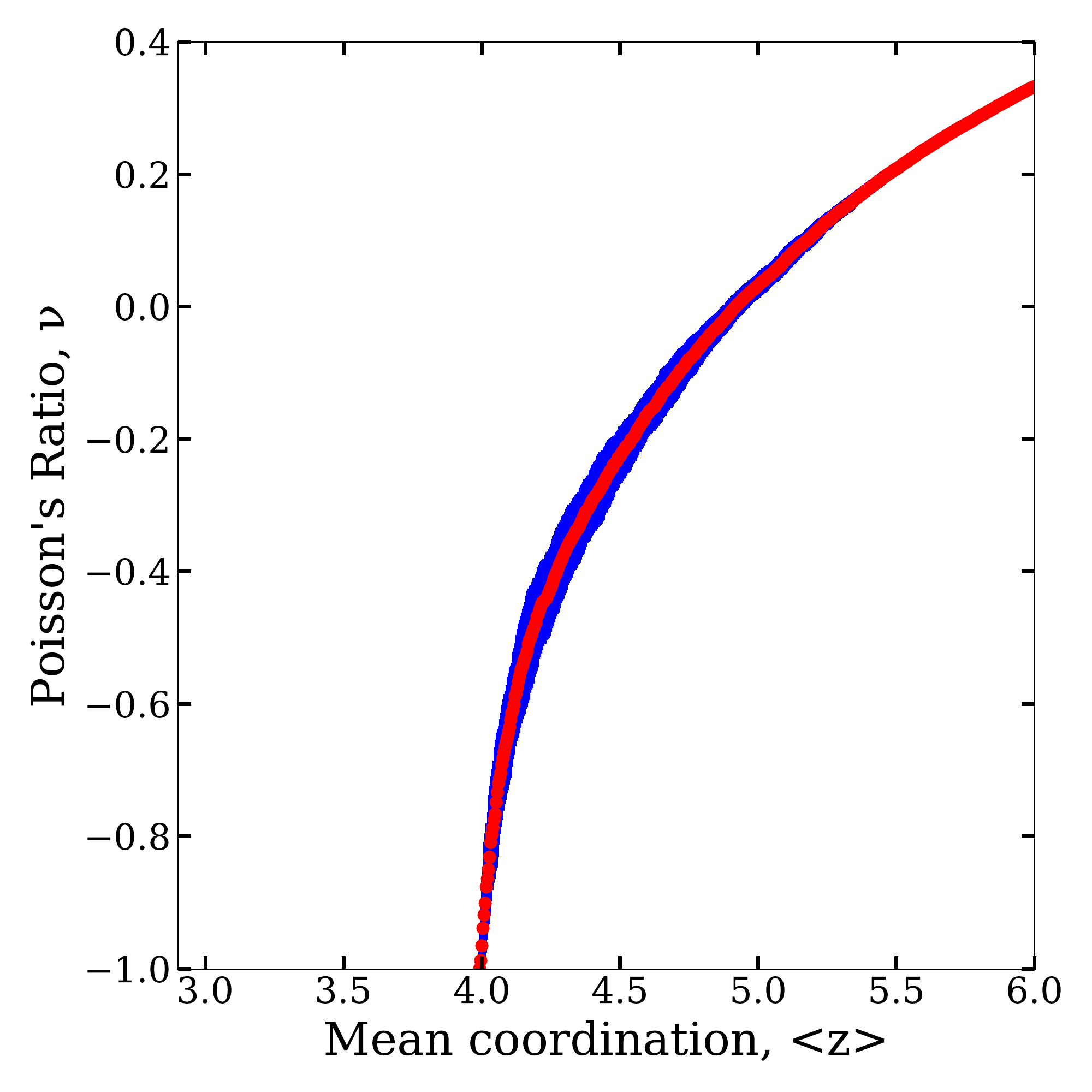}
\caption{(color online) The Poisson's ratio versus the mean coordination $\mean{z}$ for the 50 samples used in Figure~\ref{fig:plots}. The red dots along the central line show the value of the averaged Poisson's ratio and the blue vertical bars show the standard deviation for each data point.}
\label{fig:ratio}
\end{figure}

Figure~\ref{fig:ratio} shows the behavior of the ensemble-averaged Poisson's ratio as the networks are pruned. The central line, shown as red, is the Poisson's ratio, while the blue vertical bars highlight the standard deviation of the measurements over 50 samples, each with $N=500$ nodes. As can be seen from the plot, the Poisson's ratio of these systems spans over the range $(-1, 1/3)$. The small standard deviations mean that this method can be used to design and build any disordered convex structure with a desired Poisson's ratio by choosing the corresponding mean coordination that can be read from Figure~\ref{fig:ratio}. 

In Summary, here we introduce a method to produce disordered auxetic networks with near neighbor forces without re-entrant polygons in $2$D. The algorithm that we use produces networks with any desired value of the Poisson's ratio in the range $-1 < \nu < 1/3$ by tuning the mean coordination $\mean{z}$ down from $6$ to $4$ using a specific protocol. This protocol involves removing edges that minimally reduce the shear modulus while maintaining Hilbert's mechanical stability condition at each node. Any desired value of the Poisson ratio can be acheived by this method, all the way down to $-1$. Starting from a Delaunay triangulation, this leads to a disordered network where all the polygons remain convex at every stage. We chose all the spring constants to be the same, but they could differ by factors of $2$ etc., and similar results would be obtained. This result remains quite perplexing and we have no easy geometric explanation at this time Examination of Figure~\ref{fig:network} shows that while many of the polygons are far from their maximum area, none are pathologically compressed - with width/length ratios for each polygon being typically in the range $1$ to $2$. We anticipate that similar results can be obtained in $3$D.

We acknowledge useful discussions with Sidney Nagel, Andrea Liu, Louis Theran, and Mahdi Sadjadi. The work at Arizona State University is supported by the National Science Foundation under grant DMS 1564468. This work used the Extreme Science and Engineering Discovery Environment (XSEDE), which is supported by National Science Foundation grant number ACI-1548562.

\bibliography{Auxetic}
\end{document}